\documentclass[aps,prl,twocolumn,superscriptaddress]{revtex4-1}

\usepackage{amsmath}
\usepackage{amsfonts}
\usepackage{amssymb}
\usepackage{relsize}
\usepackage{graphicx}
\usepackage{subfigure}

\usepackage[caption=false]{subfig}

\begin{document}


\title{A novel regime for the dynamical Casimir effect}


\author{B. E. Ordaz--Mendoza}
\email[]{ordaz@phys.uconn.edu}
\affiliation{Department of Physics, University of Connecticut, Storrs, Connecticut 06269, USA}
\author{S. F. Yelin}
\affiliation{Department of Physics, University of Connecticut, Storrs, Connecticut 06269, USA}
\affiliation{Department of Physics, Harvard University, Cambridge, Massachusetts 02138, USA}


\date{\today}


\begin{abstract}
The dynamical Casimir effect (DCE) is the production of photons by the amplification of vacuum fluctuations. In this paper we demonstrate new resonance conditions in DCE that potentially allow the production of optical photons when the mechanical frequency is smaller than the lowest frequency of the cavity field. We consider a cavity with one mirror fixed and the other allowed to oscillate. In order to identify the region where production of photons takes place, we do a linear stability analysis and investigate the dynamic stability of the system under small fluctuations. By using a numerical solution of the Heisenberg equations of motion, the time evolution of the number of photons produced in the unstable region is studied.
\end{abstract}

\pacs{}

\maketitle

\begin{figure}
\includegraphics[width=85mm]{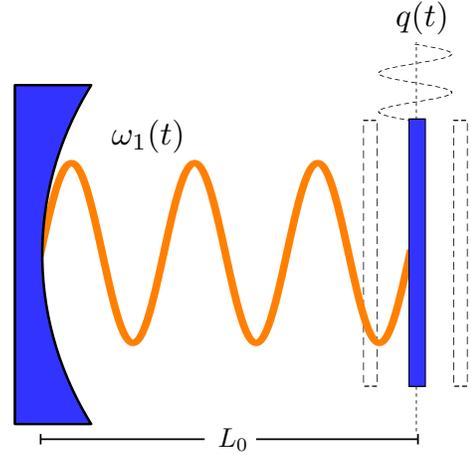}  
\caption{\label{fig1}Schematic representation of a cavity with a moving mirror. Equilibrium length and photon frequency are $L_{0}$ and $\omega_{1}(t)$, respectively. Both are periodically modified by small oscillations $q(t)$ of one of the mirrors.}  
\end{figure}

Vacuum fluctuations of the electromagnetic field are a direct manifestation of quantum effects. The dynamical Casimir effect (DCE) is the generation of photons from the quantum vacuum due to a time-dependent boundary condition of the electromagnetic field. Unlike the classical situation, the quantum vacuum contains fluctuations, and the interaction between them and the time-dependent boundary conditions can create photons. The production of photons from a quantum field confined in a cavity with time-dependent boundary conditions was first analyzed by Moore \cite{Moore1970}. A clear way to accomplish this is by changing the cavity length \cite{Moore1970,Fulling1976}, as for instance when one of the mirrors undergoes harmonic oscillations (see Fig. \ref{fig1}). The experimental demonstration of DCE at microwave frequencies was reported in \cite{WilsonDCE2011} and more recently in \cite{Lahteenmaki2013}. However, no optical frequency photons produced by DCE have been seen yet. A fundamental limitation is that the periodic modulation of the cavity imposes a definite ratio of photon-to-mechanical frequencies as a resonance condition for photon generation \cite{Dodonov1989,Dodonov1993,DCE_gen_res_97}: When the mechanical frequency $\Omega$ of the moving mirror is roughly twice the fundamental frequency $\omega_{1}$ of the unperturbed cavity, the effect of parametric resonance is the largest and the number of photons in a perfect cavity grows exponentially with time \cite{Meplan1996}; this has been the traditional resonance condition in the study of DCE. Obviously, it is hard to produce optical photons when this means that the mechanical frequencies (phonon frequencies) have to be near-UV.

The simplest description of DCE in the optical domain can be pictured as a Fabry-P\'{e}rot cavity with one mirror fixed and the other allowed to oscillate (Fig. \ref{fig1}). The coupling between the two modes takes place via the parametric modulation of the cavity by the mechanical amplitude of the mirror's motion; the photon frequency depends on the length of the resonator and therefore the photon frequency depends on time as the length of the resonator changes. The usual perturbative treatment is to assume that the maximum displacement of the mirror $x_{m}$ with respect to the unperturbed cavity length $L_{0}$ is a small parameter ($\epsilon \equiv x_{m}/L_{0} \ll 1$), and expand the photon frequency to lowest order in $\epsilon$ \cite{LawPRL1994,LawPRA1995}. For small oscillation amplitudes the linear order is sufficient while quadratic orders of $\epsilon$ are necessary for the "membrane-in-the-middle" optomechanical set up \cite{Harris2008} (in this case when the moving membrane is at an extremum of $\omega_{1}$, the linear term in $\epsilon$ vanishes and the coupling is dominantly quadratic). Higher orders of $\epsilon$ are typically neglected; nevertheless such higher orders are important as they give rise to resonance conditions in DCE that have not been explored before and that would lead to the production of optical photons.


In this paper we propose to go beyond the linear and quadratic approximations treated previously, and study the effect of higher orders of $\epsilon$ in  the resonance condition necessary for DCE. We are, in particular, interested in the regime where the mechanical frequency is smaller than the optical frequency ($\Omega < \omega_{1}$). Such a condition would help to create experimentally feasible conditions for the demonstration of DCE in the optical domain. 

Our model consists of a light field in a one-dimensional empty cavity formed by one fixed  mirror and the other moving in a time-dependent trajectory $q(t)$ prescribed externally. The maximum amplitude $x_{m}$ of the mirror's motion is small compared to the unperturbed cavity length $L_{0}$. The mirror's mechanical oscillation is treated classically and we choose its trajectory as nearly harmonic \cite{LawPRA1994}, therefore a suitable choice is \footnote{According to Moore's description of quantum field with time-dependent boundary conditions, the only possible trajectories are those that satisfy both the one-dimensional wave equation for the vector potential $\Box A(x,t) = 0$, and the boundary conditions $A(0,t) = A(x=q(t),t) =0$.}
\begin{equation}
\label{motion}
q(t) = L_{0} \exp \left[ \epsilon \sin \Omega t \right].
\end{equation} 


The cavity modes are coupled to the mechanical modes of the mirror, hence the usual opto-mechanical description of this system is based on the Hamiltonian \cite{Walls1993,*Walls1994} $H \approx \hbar \omega_{cav}(x) a^{\dagger} a + \hbar \Omega b^{\dagger} b$, where $a$ and $b$ are the annihilation operators for the light (photon) and mechanical (phonon) modes, respectively, and $x \propto (b^{\dagger} + b)$ is the displacement operator of the mirror's motion. The coupling between the two degrees of freedom is contained in the position-dependent cavity frequency $\omega_{cav}(x)$, which for small oscillation amplitudes of the mirror's motion, can be expanded around an equilibrium position $L_{0}$. In contrast, we use Law's derivation of a Lagrangian density \cite{LawPRA1994} resulting in the effective Hamiltonian 
 \begin{equation}
 \label{Heff}
 H_{\text{eff}} = \sum\limits_{\substack{k}} \omega_{k}(t) a^{\dagger}_{k}a_{k} - \frac{\dot{q}(t)}{q(t)} \, \mathlarger{\mathit{f}} ( a_{k} , a^{\dagger}_{k}, a_{j}, a^{\dagger}_{j} ),
 \end{equation}
 where
 \begin{eqnarray}
 \nonumber
 &\mathlarger{\mathit{f}}&\!\!(a_{k} , a^{\dagger}_{k}, a_{j}, a^{\dagger}_{j} ) = \frac{i}{4} \sum_{k} \left( a_{k}^{\dagger 2} - a^{2}_{k}\right) -\\
 \nonumber
 &-& \frac{i}{2} \sum\limits_{\substack{k,j \\ k\neq j}} g_{kj} \sqrt{\frac{\omega_{k}(t)}{\omega_{j}(t)}} \left( a_{k}^{\dagger}a^{\dagger}_{j} + a_{k}^{\dagger}a_{j} - a_{j}a_{k} - a^{\dagger}_{j}a_{k}\right), 
 \end{eqnarray}
 and 
 \begin{eqnarray}
 \nonumber
 g_{kj} = \begin{cases} 
 (-1)^{k + j} \frac{2 k j}{j^{2} - k^2}, & k\neq j \\
 0, & k = j.
 \end{cases}
 \end{eqnarray}
 The difference between the two is that in Eq. (\ref{Heff}) the mirror is not considered a dynamical degree of freedom, i.e., the mirror's motion is treated classically and it is a prescribed function of time. The mirror's trajectory is chosen such that $\dot{q}(t)/q(t)$ is purely harmonic, thus simplifying the calculation without making a qualitative difference.

The second term on the right hand side of Eq. (\ref{Heff}), arises due to the fact that $\omega_{k}(t)$ is an explicit function of time in the definitions of $a^{\dagger}_{k}$ and $a_{k}$ (see supplementary material), and hence the Hamiltonian that generates the equations of motion contains some additional terms. The effective Hamiltonian (\ref{Heff}) describing this system, is a time-dependent multimode Hamiltonian that contains terms associated with two-photon processes $a_{k}^{\dagger}a^{\dagger}_{j}$, and $a_{k}a_{j}$, that change the total number of photons and are responsible for DCE. The terms characterized by $a_{k}^{\dagger}a_{j}$ do not change the total number of photons.




The time-dependent photon frequencies are determined by $\omega_{k}(t) = k \pi/q(t)$, where $c=1$ and $k$ is an integer that labels each mode. Using Eq. (\ref{motion}) in the definition of the time-dependent photon frequency, and performing a Taylor series expansion in the small parameter $\epsilon$, we get
\begin{equation}\label{freqexp}
\omega_{k}(t) \approx  \omega_{k0} \left[ 1 - \epsilon \sin \Omega t + \frac{1}{2} \left( \epsilon \sin \Omega t\right)^{2} - \dots \right], 
\end{equation}
where $\omega_{k0}$ denotes the photon frequency associated with the unperturbed cavity length $L_{0}$. Since the cavity field operators depend explicitly on time through the time-dependent photon frequency $\omega_{k}(t)$, therefore by expanding the cavity field operators in $\epsilon$ and using (\ref{freqexp}), we obtain a weighted sum in terms of even and odd powers of $\epsilon$, and unperturbed annihilation and creation operators $a_{k0}$ and $a^{\dagger}_{k0}$, respectively. For the time-dependent annihilation operator of mode $k$ we get
\begin{equation}\label{aexpan}
a_{k}(t)\approx\!\!\!\sum\limits_{\substack{l=0, 2,...\\ even}}^{n}\!\!\!\frac{\left(\epsilon \sin \Omega t\right)}{2^{l} l!}^{l} a_{k0} -\!\!\! \sum\limits_{\substack{l=1, 3,\dots \\odd}}^{n} \!\!\!\frac{\left(\epsilon \sin \Omega t\right)}{2^{l} l!}^{l} a^{\dagger}_{k0},
\end{equation}
where $n$ is the maximum order of the expansion. In the rest of the article we will omit the subscript $0$ in the unperturbed photon frequency and field operators for clarity.
 

To study the influence of higher orders of $\epsilon$ in the resonance conditions necessary for DCE when $\Omega < \omega_{1}$, we start by considering $k$ modes of the cavity field in the effective Hamiltonian (\ref{Heff}), and perform a series expansion in $\epsilon$ of the photon frequency $\omega_{k}(t)$ and field operators using expressions (\ref{freqexp}) and (\ref{aexpan}), respectively. Next we obtain the Heisenberg equations of motion from the series expanded Hamiltonian and choose the resonance condition present in the system that gives the largest photon to mechanical frequency ratio. In general, such a resonance is given by
\begin{equation}\label{genres}
\tilde{\omega}_{1} = \frac{n}{2} \, \Omega, 
\end{equation} 
where $n$ is the maximum order of the series expansion of Eq.~(\ref{Heff}) in the small parameter $\epsilon$, and is chosen such that $\Omega<\omega_{1}$. Here $\tilde{\omega}_{1} \approx \omega_{1}$ is the shifted photon frequency of the fundamental mode associated with the unperturbed cavity length $L_{0}$ (see supplementary material).

Notice that the effective Hamiltonian (\ref{Heff}) is quadratic in the operators involved, therefore the equations of motion for the cavity field operators are linear in nature and constitute a closed form. A trivial steady-sate solution of this system of equations is when no photons are initially present in the cavity, i.e., $\left\langle a^{\dagger}_{i} a_{i} \right\rangle = 0$, for $i = 1,\dots, k$. This steady-state solution is stable, meaning that small perturbations from it decay in time and hence no photons are produced. In the opposite case, when this "zero-solution" is dynamically unstable, any noise (e.g., vacuum fluctuations) drives the system away from steady-state resulting in photon buildup. To determine the region where photon production takes place, we do a linear stability analysis. It is well known that a system is dynamically stable if $\operatorname{Re} \left(\lambda\right) <0$ for all eigenvalues $\lambda$ of the system.

In order to study vacuum fluctuations, a Heisenberg-Langevin formulation should be used. However, here we consider a simplified treatment where quantum fluctuations are simulated by a small deviation from the steady-state values for the initial condition of the system. 

A key to our formulation is that the series expansion of Hamiltonian (\ref{Heff}) now contains higher harmonics that result in resonance conditions that have been unexplored so far, and have the significant experimental advantage that the mechanical frequency $\Omega$ can be smaller than the unperturbed photon frequency $\omega_{1}$. For resonance to occur, notice that each slowly varying field operator in the equations of motion carries a fast oscillatory phase factor that would have to be approximately matched by the amplitude multiplying it in order to considerably contribute to the dynamics of the system. The main step in our analytical treatment is thus to neglect all the strongly non-resonant terms, i.e., the generalization of the so-called "counter-rotating" terms. With this approximation, the remaining equations of motion simplify significantly (see supplementary material) and the eigenvalue with the largest real part is given by
\begin{equation}\label{geneigenval}
\lambda_{\rm max}=\frac{1}{2\,n!} \epsilon^{n} \omega_{1} - \frac{1}{2}\kappa_{1},
\end{equation}
where $\kappa_{1}$ is the damping rate of the fundamental mode of cavity field, and we have considered the same damping rate in the rest of the cavity field modes.


\begin{figure}[hbt!]
	\includegraphics[width=84mm]{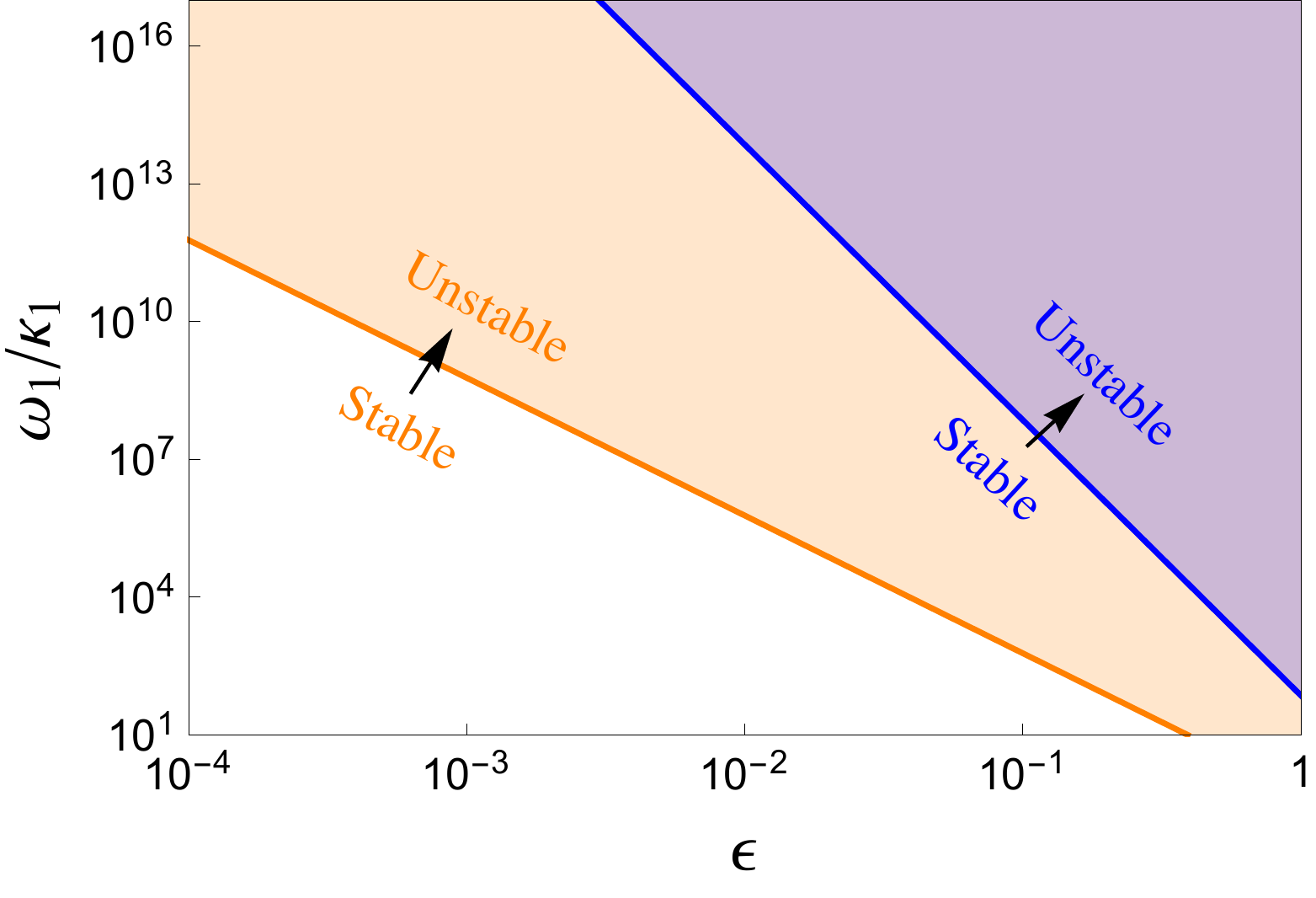}  
	\caption{\label{fig2}(Color online) Dimensionless frequency vs small parameter $\epsilon$ displaying stability of the vacuum state. Boundary of $\lambda_{\rm max}$ when expanding up to terms in $\epsilon^{3}$ (orange/lower solid line) and up to terms in $\epsilon^{6}$ (blue/upper solid line). The resonance condition is $\tilde{\omega}_{1} = 3\,\Omega/2$ and $\tilde{\omega}_{1} = 3 \, \Omega$, respectively. The number of photons grows faster the deeper one is in the unstable parameter regime.}  
\end{figure}
As an example, consider a series expansion of (\ref{Heff}) up to $\epsilon^{3}$ terms and the lowest three modes of the cavity field. The equations of motion for the cavity field operators in the presence of photon damping result in a closed form, and we choose the resonance condition such that $\Omega<\omega_{1}$, in this case with $n = 3$ in (\ref{genres}) the resonance condition is $\tilde{\omega}_{1} = 3\,\Omega/2$. 
In order to determine the region of operation for DCE, notice that $\lambda_{\rm max} = 0$ describes the boundary where the maximum eigenvalue changes sign and therefore the system transitions from stable to unstable. Here the maximum eigenvalue is given according to Eq. (\ref{geneigenval}) with $n$ substituted by 3. 
In Fig. \ref{fig2} we plot the ratio of the unperturbed photon frequency to the photon damping rate as a function of $\epsilon$ (orange/lower solid line). For example, when $\omega_{1}/\kappa_{1} = 1\times10^{3}$ and $\epsilon = 0.45$, the system operates in the unstable regime (orange region) and production of photons by the parametric modulation of the cavity mirror takes place.


\begin{figure}[h!]
	\includegraphics[width=84mm]{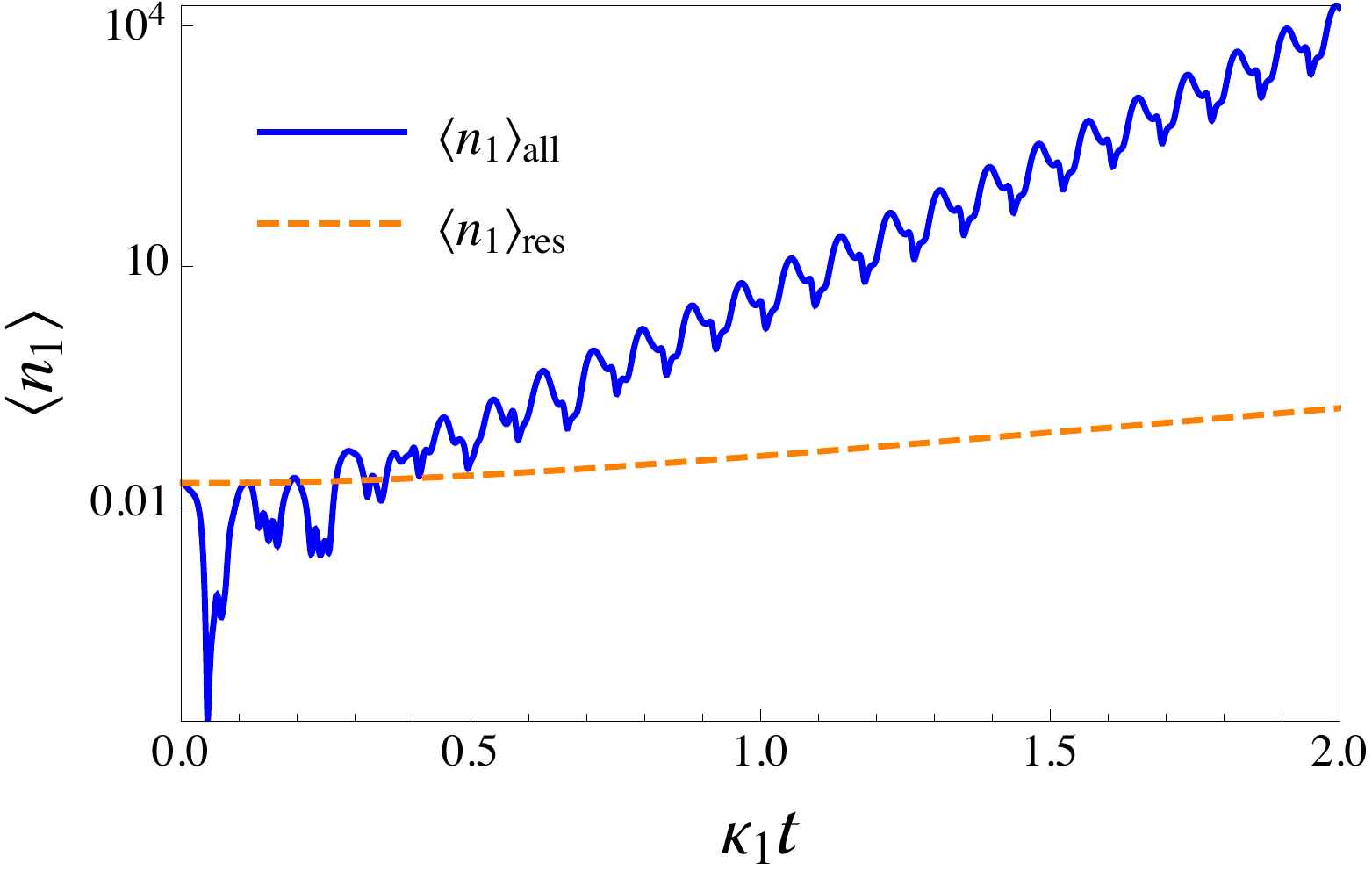}
	\caption{\label{fig3}Time evolution of the mean photon number in the lowest mode vs dimensionless time. Numerical solution of the Heisenberg equations of motion when including resonant and nonresonant terms (blue, solid line), and resonant terms only (orange, dashed line). With $\omega_{1}/\kappa_{1}=1\times10^{3}$ and $\epsilon=.45$ the system operates in the unstable regime.}
\end{figure}

Having identified the region of operation of our model, now it is interesting to determine how fast photon generation occurs especially if one does not neglect the fast-oscillating terms. Therefore, we solve numerically the Heisenberg equations of motion using the effective Hamiltonian (\ref{Heff}) for $k$ modes of the field and expanding up to terms in $\epsilon^{n}$. By looking at the equations of motion and using the resonant condition (\ref{genres}), we note that the contribution of the terms for higher order than $n$ in $\epsilon$ is negligibly small, therefore we cut off any higher order.  

In Fig. \ref{fig3} we plot the full time evolution of the mean number of photons produced in the lowest cavity mode $\left\langle  n_{1}\right\rangle$ when expanding the effective Hamiltonian (\ref{Heff}) up to terms in $\epsilon^{3}$ and considering the lowest three modes of the cavity field. From Eq. (\ref{genres}) the resonance condition with $n=3$ is $\tilde{\omega}_{1} = 3\,\Omega/2$, we take $\omega_{1}/\kappa_{1}=1\times10^{3}$ and $\epsilon=.45$ such that the system operates in the unstable regime, and resonant production of photons from vacuum fluctuations occurs. Notice that the solution for the number of photons grows faster when including resonant and nonresonant terms in the equations of motion (blue, solid line), than when including resonant terms only (orange, dashed line). Qualitatively the dynamics -- in particular, the onset of photon production -- is well described by the simplified analytical treatment. Quantitatively this is not exact and the behavior needs the full quantum description. The simplified treatment gives an idea of the region of DCE at optical frequencies when $\Omega < \omega_{1}$ and can be solved analytically.

\begin{figure}[h!]
	\includegraphics[width=84mm]{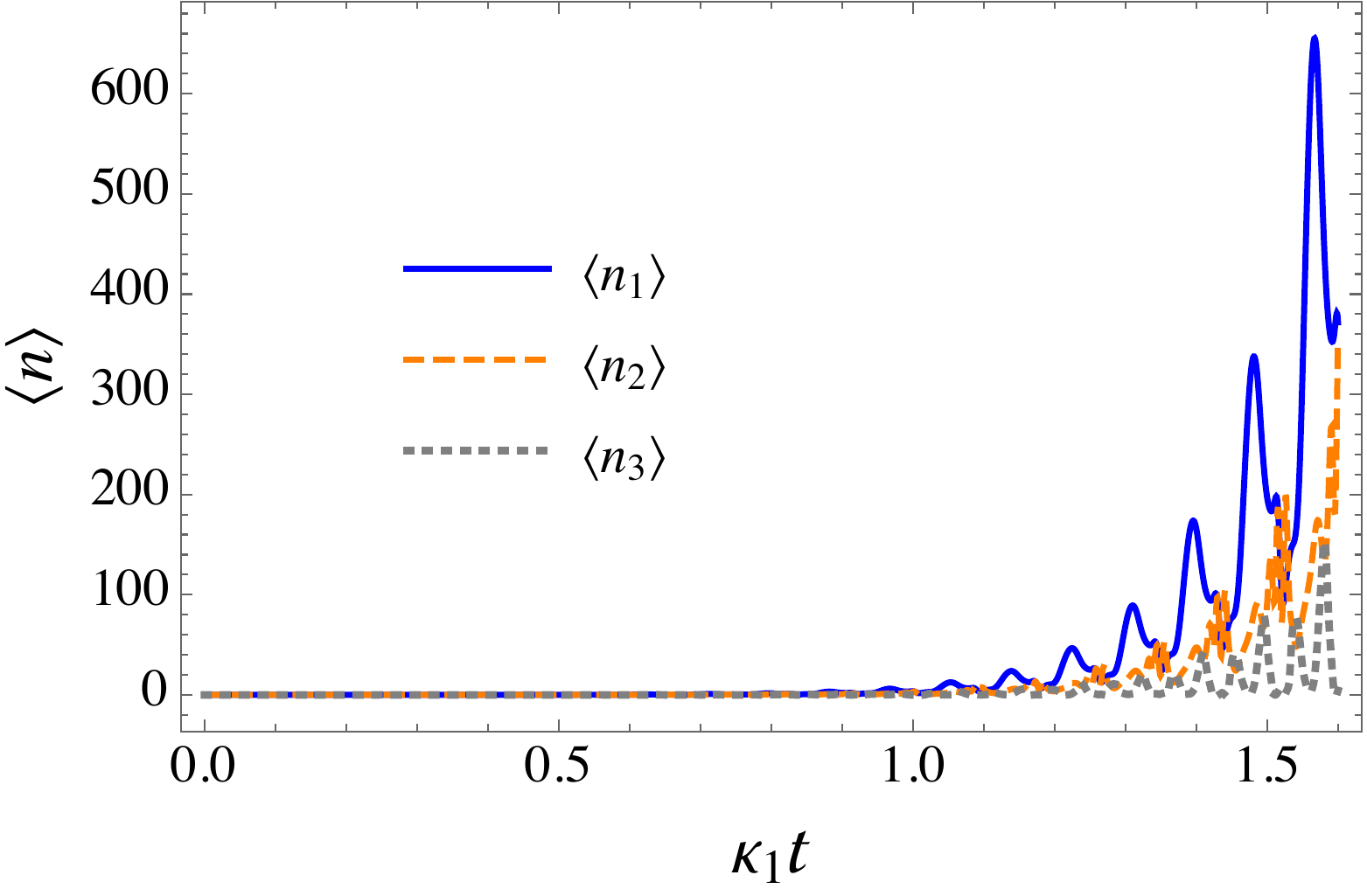}  
	\caption{\label{fig4} Time evolution of the mean photon number in modes $1$, $2$ and $3$ vs dimensionless time. When the system operates in the unstable regime and considering resonant and nonresonant terms in the equations of motion, the number of photons of the scattered modes $2$ and $3$ also grows exponentially on the average at a slower rate. The parameters are the same as in the previous figure.}  
\end{figure}

The effective Hamiltonian (\ref{Heff}) contains terms associated with the scattering of photons from mode $k$ to mode $j$, hence it is relevant to study the photon evolution of other modes than the fundamental one. In Fig. \ref{fig4} we plot the time evolution of the average number of resonant photons in the first three modes of the cavity field when considering terms up to $\epsilon^{3}$ in the expanded Hamiltonian. Starting slightly away from the steady-state, with $\omega_{1}/\kappa_{1}=1\times10^{3}$ and $\epsilon=.45$, the system operates in the unstable regime and DCE occurs. Using the resonant condition $\tilde{\omega}_{1} = 3 \,\Omega/2$, the number of photons in all modes considered is parametrically excited. The number of scattered photons from mode $1$ to $2$ and from mode $1$ to $3$ also grow exponentially in time at a slower rate.   

\begin{figure}[h!]
	\includegraphics[width=84mm]{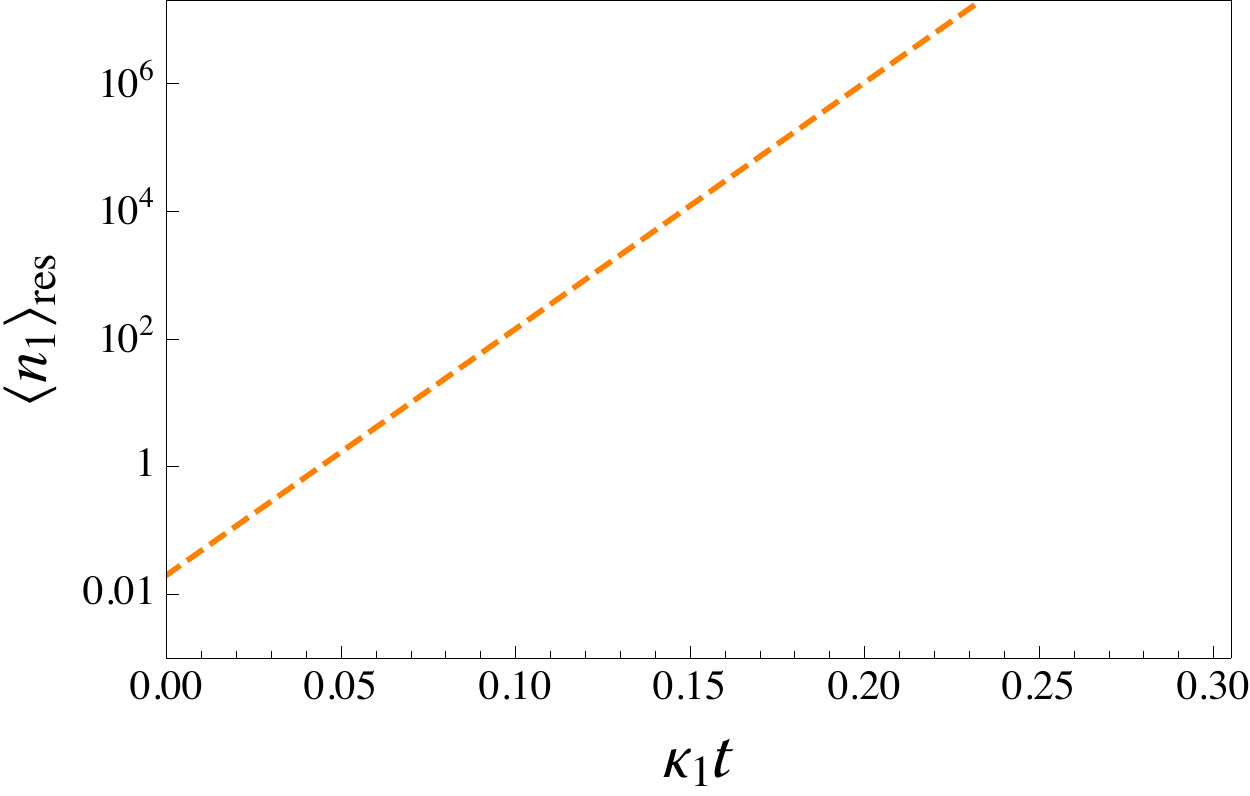}
	\caption{\label{fig5} Time evolution of the mean photon number in the fundamental mode vs dimensionless time. Numerical solution of the Heisenberg equations of motion when considering resonant terms only and expanding up to $\epsilon^{6}$ terms. With $\omega_{1}/\kappa_{1}=1\times10^{16}$ and $\epsilon=.02$ the system operates in the unstable regime and DCE at optical frequencies takes place.}
\end{figure}

One of the interesting questions is now how to reach even higher photon frequencies, e.g., optical frequencies. With the same treatment for $n = 6$, $k=6$ the general resonant condition Eq. (\ref{genres}) in this case is $\tilde{\omega}_{1} = 3 \, \Omega$. In Fig. \ref{fig2} we represent the boundary when $\lambda_{\rm max}$ changes sign (blue/upper solid line). For example, when $\omega_{1}/\kappa_{1} = 1 \times 10^{16}$ and $\epsilon = .02$, the system operates in the unstable regime (blue shaded region). Here, however, we did not conclude the calculation of the quantitatively correct full set of resonant and non-resonant terms. The reason is that this is simply numerically very unstable because of the extremely fast oscillations. The qualitative limit, however, shows correctly the order of magnitude of the onset of photon production (see Fig. \ref{fig5}).

To conclude, we have shown that the presence of higher orders of the small parameter $\epsilon$ in the expansions of the photon frequency and field operators, gives rise to resonances in the DCE that have not been explored before. This constitutes a novel regime of operation of the DCE at potentially higher frequencies when the mechanical frequency of the moving mirror is smaller than the photon frequency. Different from the usual resonance condition studied in DCE that involves moving a mirror at higher than light frequencies, our region of operation potentially permits resonances at mechanical frequencies that are experimentally feasible for its demonstration in the optical domain. Another way that is potentially more promising, is the equivalent change of the absorptive or refractive properties of the cavity, e.g., by placing and manipulating radiators into the cavity \cite{Yablo1989,DCE_indexref,MIR_DCE2005}. 

The small parameter $\epsilon$ is related to the single-photon coupling strength for optomechanical set-ups allowing a linear coupling (cavity with a moving mirror) as $g^{(1)}_{0}=\epsilon \,\omega_{1}$, and for quadratic coupling as $g^{(2)}_{0}=\epsilon^{2}\, \omega_{1}/2$; experiments exhibiting this type of coupling include: "membrane-in-the-middle" systems \cite{Harris2008,Harris2010,APLHarris2012}, trapped cold atoms \cite{TrapatomsPRL2010}, trapped microspheres \cite{SphereNature2011} or double-disk structures \cite{PainterPRL2009}. The upper limit of the small parameter $\epsilon$ can be determined as follows: $\epsilon = \frac{x_m}{L_0} = \frac{v_m}{\Omega L_0} < \frac{c}{\Omega L_0} \equiv \epsilon_{\rm max}$  if there is no principal limit of the amount of phonons, i.e., the amplitude of the mechanical drive. For example, for THz phonons and a cavity length of $L_{0} = 3$ cm, this would result in a limit of about $\epsilon \approx 0.01$. For experimental parameters, like in \cite{Harris2008}, with a membrane of effective mass $m_{\text{eff}}=43$ ngr, mechanical frequency $\Omega/2 \pi = 354.6$ kHz and for a cavity length of $L_{0} \approx 6$ cm, on the other hand, $\frac{c}{\Omega L_0} \approx 10^4$, that is, for mirror movement at the speed of light the mirror would be displaced way beyond the size of the cavity, and thus there is no relativistic principal limit on epsilon, which is now only limited by the accessible driving intensity.

As an outlook to improve the effective coupling quantity $\epsilon$ further, one could envision ``coupling amplification'' using additional sources, e.g, like in \cite{Clerk2016}, i.e., introducing a strongly detuned parametric drive to the mechanics. Finally, for a situation with much smaller mechanical motion, this motion should be quantized into phonons. In this -- for optomechanical systems typically -- situation, it is particularly interesting to treat the number of phonons also as a dynamical variable. In the strong coupling regime, this poses the question what the nonlinear dynamics of the type investigated in this letter will affect. This will be discussed in an upcoming article.

\begin{acknowledgments}
B.E.O.M acknowledges financial support from CONACYT, Mexico. S.F.Y. wishes to thank the National Science Foundation and the Army Research Office for funding.  
\end{acknowledgments}

\bibliographystyle{apsrev4-1}
\bibliography{mybib}

\pagebreak
\widetext
\begin{center}
	\textbf{\large Supplementary Material for: A novel regime for the dynamical Casimir effect}
\end{center}
\setcounter{equation}{0}
\setcounter{figure}{0}
\setcounter{table}{0}
\setcounter{page}{1}
\makeatletter
\renewcommand{\theequation}{S\arabic{equation}}
\renewcommand{\thefigure}{S\arabic{figure}}
\renewcommand{\bibnumfmt}[1]{[S#1]}
\renewcommand{\citenumfont}[1]{S#1}

\section{Derivation of the effective Hamiltonian}\label{sec1}
The derivation of the effective Hamiltonian is presented with great detail in [12], here we mention a few remarks. Considering the time-dependent creation and annihilation operators for cavity mode $k$. 
\begin{equation*}
a_k(t) = \sqrt{\frac{1}{2 \omega_{k}(t)}}\left[\omega_{k}(t) Q_{k} + i P_{k} \right], \qquad a^{\dagger}_{k}(t) = \sqrt{\frac{1}{2 \omega_{k}(t)}}\left[\omega_{k}(t) Q_{k} - i P_{k} \right], 
\end{equation*}
where $\hslash = 1$ and $\omega_{k}(t)$ is the time-dependent cavity frequency of mode $k$. Here $Q_{k}$ and $P_{k}$ are the generalized position and momentum operators for the cavity field, respectively, which obey the commutation relations $[Q_{k}, P_{j}] = i \delta_{kj}$ which guarantee that $[a_{k}(t), a^{\dagger}_{j}(t) ] = \delta_{kj}$.

The Lagrangian density of the system is given by
\begin{equation}
L(x,t) = \frac{1}{2} \left[ \left( \frac{\partial A(x,t)}{\partial t}\right)^{2} - \left( \frac{\partial A(x,t)}{\partial x}\right)^{2} \right], 
\end{equation}
where $c=1$ and $A(x,t)$ is the vector potential. The quantization is achieved by letting the vector potential be a quantum operator. Expanding in terms of an orthonormal set of mode functions $\phi_{k}(x;t)$ [12] 
\begin{equation}\label{S_expVecPot}
A(x,t) = \sum_{k} Q_{k}(t) \phi_{k}(x;t).
\end{equation}
The conjugate momentum is obtained by
\begin{equation}
\pi(x,t) = \dfrac{\partial A(x,t)}{\partial t},
\end{equation}
and expanding in terms of $\phi_{k}(x;t)$
\begin{equation}\label{S_expMomentum}
\pi(x,t) = \sum_{k} P_{k}(t) \phi_{k}(x;t).
\end{equation}
By inverting Eqs. (\ref{S_expMomentum}) and (\ref{S_expMomentum}), and taking their time derivative, one obtains the equations of motion for $Q(t)$ and $P(t)$ and then the Hamiltonian that generates such equations of motion is constructed. For the annihilation and creation operators the equations of motion are obtained in the same way, hence with the explicit form of the mode functions  
\begin{equation}
\phi_{k}(x;t) = \sqrt{\frac{2}{q(t)}} \sin \frac{k \pi x}{q(t)}, 
\end{equation}
the effective Hamiltonian (Eq. (2) on the main text) follows.

\section{Shifted cavity frequency}\label{sec2}

The shifted cavity frequency in general is given by
\begin{equation}\label{S_shiftfreq}
\tilde{\omega}_{1} = \omega_{1} \cosh^{2} \left( \epsilon \left\langle \sin \Omega t\right\rangle \right), 
\end{equation}
where $\left\langle \, \right\rangle $ is the average value of a function. Then performing a series expansion of Eq. (\ref{S_shiftfreq}) up to $n$
\begin{equation*}
\tilde{\omega}_{1} \approx \frac{1}{2} \left[1+ \sum\limits_{l=0}^{n} \frac{4^{l}}{(2 l)!} \epsilon^{2 l} \left\langle \sin^{2 l} \Omega t \right\rangle \right] \omega_{1}, 
\end{equation*}
since
\begin{equation*}
\left\langle \sin^{2 l} \Omega t\right\rangle = \frac{1}{4^{l} } \binom{2 l}{l}, \quad \text{for} \quad l= 0,1,2,\dots, 
\end{equation*}
then the shifted cavity frequency is determined by 
\begin{equation}\label{S_genshiftfreq}
\tilde{\omega}_{1} \approx \frac{1}{2} \left[1+ \sum\limits_{l=0}^{n} \frac{1}{(l!)^{2}}\epsilon^{2 l} \right] \omega_{1}. 
\end{equation}
We use Eq. (\ref{S_genshiftfreq}) with $n=3$ and $n=6$ for the shifted cavity frequency in the main text.

\section{Example: equations of motion for $k=3$ and $n=3$}\label{sec3}
By using the series expansions up to terms in $\epsilon^{3}$ of the cavity frequency and field operators Eqs. (3) and (4), respectively (all of the equations here refer to the main text), and taking the lowest 3 modes of the cavity field, one gets a series expansion of the effective Hamiltonian (2). Then using the mirror's trajectory Eq. (1), the Heisenberg equations of motion in a frame rotating at a frequency $\tilde{\omega}_{1} + \tilde{\omega}_{2} + \tilde{\omega}_{3}$ are

\begin{eqnarray}
\nonumber
\frac{d a_{1}}{dt} &=& - \frac{\kappa_{1}}{2} a_{1} + \epsilon \left[ \frac{\omega_{1}}{2} \left( e^{i \Omega t} - e^{-i \Omega t} \right) \left( a_{1} + a^{\dagger}_{1} e^{2 i \tilde{\omega}_{1} t} \right) - \frac{\Omega}{4} \left( e^{i \Omega t} + e^{-i \Omega t} \right) a^{\dagger}_{1} e^{2 i \tilde{\omega}_{1} t} \right. -\\
\nonumber
&-& \left. \frac{\Omega}{2} \left( \frac{\sqrt{\omega_{1} \omega_{2}}}{\omega_{2} - \omega_{1}}\right) \left( e^{i \Omega t} + e^{-i \Omega t} \right) a_{2} e^{- i (\tilde{\omega}_{2} - \tilde{\omega}_{1} ) t} + \frac{\Omega}{2} \left( \frac{\sqrt{\omega_{1} \omega_{2}}}{\omega_{2} + \omega_{1}}\right) \left( e^{i \Omega t} + e^{-i \Omega t} \right) a^{\dagger}_{2} e^{i (\tilde{\omega}_{2} + \tilde{\omega}_{1}) t} \right. +\\
\nonumber
&+& \left. \frac{\Omega}{2} \left( \frac{\sqrt{\omega_{1} \omega_{3}}}{\omega_{3} - \omega_{1}}\right) \left( e^{i \Omega t} + e^{-i \Omega t} \right) a_{3} e^{- i (\tilde{\omega}_{3} - \tilde{\omega}_{1} ) t} - \frac{\Omega}{2} \left( \frac{\sqrt{\omega_{1} \omega_{3}}}{\omega_{3} + \omega_{1}}\right) \left( e^{i \Omega t} + e^{-i \Omega t} \right) a^{\dagger}_{3} e^{i (\tilde{\omega}_{3} + \tilde{\omega}_{1}) t} \right] +\\
\nonumber
&+& \epsilon^{2} \left[ \frac{i}{4} \omega_{1} \left( e^{2 i \Omega t} + e^{- 2 i \Omega t} \right) \left( a_{1} + a_{1}^{\dagger} e^{2 i \tilde{\omega}_{1} t} \right) - \frac{i}{2} \omega_{1} a_{1}^{\dagger} e^{2 i \tilde{\omega}_{1} t} \right] +\\
\nonumber
&+& \epsilon^{3} \left[ \frac{\omega_{1}}{4} \left( e^{i \Omega t} - e^{-i \Omega t} \right) \left( a_{1} + a^{\dagger}_{1} e^{2 i \tilde{\omega}_{1} t} \right) - \frac{\omega_{1}}{12} \left( e^{3 i \Omega t} - e^{-3 i \Omega t} \right) \left( a_{1} + a^{\dagger}_{1} e^{2 i \tilde{\omega}_{1} t} \right) \right],
\end{eqnarray}

\begin{eqnarray}
\nonumber
\frac{d a_{2}}{dt} &=& - \frac{\kappa_{2}}{2} a_{2} + \epsilon \left[ \frac{\omega_{2}}{2} \left( e^{i \Omega t} - e^{-i \Omega t} \right) \left( a_{2} + a^{\dagger}_{2} e^{2 i \tilde{\omega}_{2} t} \right) - \frac{\Omega}{4} \left( e^{i \Omega t} + e^{-i \Omega t} \right) a^{\dagger}_{2} e^{2 i \tilde{\omega}_{2} t} \right. +\\
\nonumber
&+& \left. \frac{\Omega}{2} \left( \frac{\sqrt{\omega_{1} \omega_{2}}}{\omega_{2} - \omega_{1}}\right) \left( e^{i \Omega t} + e^{-i \Omega t} \right) a_{1} e^{i (\tilde{\omega}_{2} - \tilde{\omega}_{1} ) t} + \frac{\Omega}{2} \left( \frac{\sqrt{\omega_{1} \omega_{2}}}{\omega_{2} + \omega_{1}}\right) \left( e^{i \Omega t} + e^{-i \Omega t} \right) a^{\dagger}_{1} e^{i (\tilde{\omega}_{2} + \tilde{\omega}_{1}) t} \right. -\\
\nonumber
&-& \left. \frac{\Omega}{2} \left( \frac{\sqrt{\omega_{2} \omega_{3}}}{\omega_{3} - \omega_{2}}\right) \left( e^{i \Omega t} + e^{-i \Omega t} \right) a_{3} e^{- i (\tilde{\omega}_{3} - \tilde{\omega}_{2} ) t} + \frac{\Omega}{2} \left( \frac{\sqrt{\omega_{2} \omega_{3}}}{\omega_{3} + \omega_{2}}\right) \left( e^{i \Omega t} + e^{-i \Omega t} \right) a^{\dagger}_{3} e^{i (\tilde{\omega}_{3} + \tilde{\omega}_{2}) t} \right] +\\
\nonumber
&+& \epsilon^{2} \left[ \frac{i}{4} \omega_{2} \left( e^{2 i \Omega t} + e^{- 2 i \Omega t} \right) \left( a_{2} + a^{\dagger}_{2} e^{2 i \tilde{\omega}_{2} t} \right) - \frac{i}{2} \omega_{2} a_{2}^{\dagger} e^{2 i \tilde{\omega}_{2} t} \right] +\\
\nonumber
&+& \epsilon^{3} \left[ \frac{\omega_{2}}{4} \left( e^{i \Omega t} - e^{-i \Omega t} \right) \left( a_{2} + a^{\dagger}_{2} e^{2 i \tilde{\omega}_{2} t} \right) - \frac{\omega_{2}}{12} \left( e^{3 i \Omega t} - e^{-3 i \Omega t} \right) \left( a_{2} + a^{\dagger}_{2} e^{2 i \tilde{\omega}_{2} t} \right) \right],
\end{eqnarray}

\begin{eqnarray}
\nonumber
\frac{d a_{3}}{dt} &=& - \frac{\kappa_{3}}{2} a_{3} + \epsilon \left[ \frac{\omega_{3}}{2} \left( e^{i \Omega t} - e^{-i \Omega t} \right) \left( a_{3} + a^{\dagger}_{3} e^{2 i \tilde{\omega}_{3} t} \right) - \frac{\Omega}{4} \left( e^{i \Omega t} + e^{-i \Omega t} \right) a^{\dagger}_{3} e^{2 i \tilde{\omega}_{3} t} \right. -\\
\nonumber
&-& \left. \frac{\Omega}{2} \left( \frac{\sqrt{\omega_{1} \omega_{3}}}{\omega_{3} - \omega_{1}}\right) \left( e^{i \Omega t} + e^{-i \Omega t} \right) a_{1} e^{i (\tilde{\omega}_{3} - \tilde{\omega}_{1} ) t} - \frac{\Omega}{2} \left( \frac{\sqrt{\omega_{1} \omega_{3}}}{\omega_{3} + \omega_{1}}\right) \left( e^{i \Omega t} + e^{-i \Omega t} \right) a^{\dagger}_{1} e^{i (\tilde{\omega}_{3} + \tilde{\omega}_{1}) t} \right. +\\
\nonumber
&+& \left. \frac{\Omega}{2} \left( \frac{\sqrt{\omega_{2} \omega_{3}}}{\omega_{3} - \omega_{2}}\right) \left( e^{i \Omega t} + e^{-i \Omega t} \right) a_{2} e^{i (\tilde{\omega}_{3} - \tilde{\omega}_{2} ) t} + \frac{\Omega}{2} \left( \frac{\sqrt{\omega_{2} \omega_{3}}}{\omega_{3} + \omega_{2}}\right) \left( e^{i \Omega t} + e^{-i \Omega t} \right) a^{\dagger}_{2} e^{i (\tilde{\omega}_{3} + \tilde{\omega}_{2}) t} \right] +\\
\nonumber
&+& \epsilon^{2} \left[ \frac{i}{4} \omega_{3} \left( e^{2 i \Omega t} + e^{- 2 i \Omega t} \right) \left( a_{3} + a^{\dagger}_{3} e^{2 i \tilde{\omega}_{3} t} \right) - \frac{i}{2} \omega_{3} a_{3}^{\dagger} e^{2 i \tilde{\omega}_{3} t} \right] +\\
\nonumber
&+& \epsilon^{3} \left[ \frac{\omega_{3}}{4} \left( e^{i \Omega t} - e^{-i \Omega t} \right) \left( a_{3} + a^{\dagger}_{3} e^{2 i \tilde{\omega}_{3} t} \right) - \frac{\omega_{3}}{12} \left( e^{3 i \Omega t} - e^{-3 i \Omega t} \right) \left( a_{3} + a^{\dagger}_{3} e^{2 i \tilde{\omega}_{3} t} \right) \right].
\end{eqnarray}

The equations above with their corresponding adjoints constitute a closed form. The steady-state of such systems of equations has the trivial solution 
$\left\langle a_{i}\right\rangle_{\text{s.s}} = 0$, for $i = 1,2,3$.  

Since the time-dependent cavity frequency is defined by $\omega_{k}(t) = k \pi/q(t)$, where $c=1$, hence the spectrum is equally spaced and $\omega_{2} = 2 \, \omega_{1}$, $\omega_{3} = 3 \, \omega_{1}$, similarly $\tilde{\omega}_{2} = 2 \, \tilde{\omega}_{1}$, $\tilde{\omega}_{3} = 3\,\tilde{\omega}_{1}$, etc. By using the resonant condition $\tilde{\omega}_{1}= 3\,\Omega/2$ in the equations of motion above, the terms that contribute to the dynamics are those that are in resonance, therefore the equations of motion simplify substantially when one neglects all strongly non-resonant terms, and the resonant equations of motion are
\begin{equation}
\frac{d}{dt} \begin{pmatrix} 
\left\langle a_{1}\right\rangle  \\ 
\left\langle a_{2}\right\rangle \\  
\left\langle a_{3}\right\rangle \\  
\left\langle a_{1}\right\rangle^{*} \\  
\left\langle a_{2}\right\rangle^{*} \\  
\left\langle a_{3}\right\rangle^{*} \\  
\end{pmatrix} =	\begin{pmatrix}
- \frac{\kappa_{1}}{2} & 0& 0& \frac{1}{12} \epsilon^{3} \omega_{1}& 0 &  0&    \\
0& - \frac{\kappa_{2}}{2} & 0& 0& 0&   0&    \\
0&0& - \frac{\kappa_{3}}{2} & 0& 0&   0&   \\
\frac{1}{12} \epsilon^{3} \omega_{1}& 0& 0& - \frac{\kappa_{1}}{2} & 0 &  0&   \\
0& 0& 0& 0 & - \frac{\kappa_{2}}{2} & 0&  \\
0& 0 & 0& 0& 0& - \frac{\kappa_{3}}{2} & 
\end{pmatrix} \begin{pmatrix} 
\left\langle a_{1}\right\rangle  \\ 
\left\langle a_{2}\right\rangle \\  
\left\langle a_{3}\right\rangle \\  
\left\langle a_{1}\right\rangle^{*} \\  
\left\langle a_{2}\right\rangle^{*} \\  
\left\langle a_{3}\right\rangle^{*}.
\end{pmatrix}
\end{equation}
We find the characteristic polynomial of the matrix above, and look for the largest real part of the eigenvalues, which is given by Eq. (6) in the main text with $n=3$.


\end{document}